\documentstyle[epsf]{article}
\begin{document}

\title
{\Large \bf      Search for Higgs Boson at LHC \\
     in the Reaction $pp\to\gamma\gamma+jet$ at a Low Luminosity
     \thanks{To appear in: {\it Proceedings of XIIth International 
     Workshop: High Energy Physics and Quantum Field Theory (QFTHEP-97)},  
     Samara, Russia, September 4-10, 1997.
     Preliminary version of this paper was included in \cite{CMSournote}, 
     where we studied also the possibilities of Higgs signal observation 
     in the final state $\gamma\gamma+lepton$ for the LHC
     high luminosity regime. }
}

\author{M.N.~Dubinin, V.A.~Ilyin and V.I.~Savrin\\
           {\small \it Institute of Nuclear Physics, Moscow State
                       University, 119899 Moscow, Russia}
       }

\date{}

\maketitle

\begin{abstract}

We discuss the SM Higgs discovery potential of the LHC in the channel $pp\to H+
jet\to\gamma\gamma+jet$ when the jet is observed  at sufficiently high $p_t$
and a small rapidity to be reliably identified. We calculate all the signal
subprocesses and the irreducible background with realistic kinematical cuts.
The reducible QCD background is also estimated.

We conclude that the channel $\gamma\gamma+jet$ can give about 120-200 signal
events for Higgs mass $M_H=$100-140 GeV  at the integrated luminosity of
30fb$^{-1}$. This signal rate should be compared with only $\sim$330-600 events
for the irreducible background  per two-photon invariant mass interval of
2~GeV. We estimate the QCD reducible background at the level of $\le 20$\% of
the irreducible one. Thus, one may hope that the Higgs  boson can be discovered
already during the LHC operation at a low luminosity. At a high luminosity of
10$^{34}$ cm$^{-2}$s$^{-1}$ the observation of several hundreds of high $p_t$ Higgs
bosons in this channel will be possible with significance higher than 15 for
$L \sim$ 100fb$^{-1}$.

\end{abstract}

%===============================================================
\section{Introduction}

It is well known that the observation of a light Higgs boson ($M_H<140$ GeV) at
the LHC collider in the inclusive channel $pp\to\gamma\gamma+X$ is not easy
\cite{inclusive,cmsTP,atlasTDR}. Even if we can reduce the misidentification of
two jets as photons to a small level,  the irreducible background from  the
$q\bar q\to\gamma\gamma$ and $gg\to\gamma\gamma$  subprocesses rapidly
increases for smaller $\gamma\gamma$ pair invariant masses, and it is necessary
to separate a rather elusive Higgs boson signal from it.  In this situation it
is important to understand whether we can  observe at the LHC any other Higgs
boson production mechanisms. In this paper we are considering the Higgs
discovery potential in the channel $pp\to\gamma\gamma+jet$ when the jet  is
observed at a sufficiently high transverse momentum and small rapidity to be
reliably identified. In this channel the signal cross section is much smaller
in comparison with the inclusive $pp\to\gamma\gamma+X$ case, but at the same
time the situation with the background is undoubtedly much better. Transition
to larger $p_t$ allows us to reduce the overwhelming backgrounds of the
completely inclusive channel mentioned above. We consider the scalar boson
production in the framework of Standard Model (SM) with one Higgs doublet.

The idea to look for the Higgs boson signal associated with a high $p_t$ jet in
the final state was considered in \cite{j1}, where the matrix elements of
signal subprocesses $gg\to g+H$, $gq\to q+H$ and $q\bar q\to g+H$ were
calculated  analytically in the  leading order $\alpha_s^3$. In \cite{Kao} the
corresponding subprocesses were calculated for the case of CP-odd Higgs boson
production within the minimal supersymmetric extension of the Standard Model
(MSSM). Recently (see Ref. \cite{alan}) the matrix elements for subprocesses
$gq\to H^\pm q'\,,H^0_i q$ were calculated analytically for charged and neutral
scalars in the framework of MSSM. However, in \cite{Kao,alan} decay channels of
the Higgs were not considered and in \cite{j1} only the final state
$\tau^+\tau^- +jet$ was discussed. In \cite{j2} the SM subprocesses were
considered for the case of  heavy Higgs boson decaying into the $WW$ or $ZZ$
pairs. Very promising numbers have been obtained recently in \cite{RainZepp}
for $\gamma\gamma+2jets$ final state with two very forward jets
($|\eta_{jet}|<5$), when Higgs scalars are produced by weak boson fusion
mechanism. In this reaction 10-20 Higgs signal events could be observed with
the significance equal to 3.5-7 at the integrated luminosity 10 fb$^{-1}$. In
\cite{abdullin} (see also \cite{cmsTP}) the final state $\gamma\gamma+(\geq
2jets)$ was simulated by means of PYTHIA generator \cite{pythia} in the
realistic CMS detector environment. The events with relatively small rapidity
jets and the events with very forward jets, produced by weak boson fusion
mechanism, were studied there (as we mentioned above, recent independent
analysis of the fusion can be found in \cite{RainZepp}). It was shown that
this signature has good prospects for the light Higgs boson search giving
$\sim$100-150 signal events at the integrated luminosity of 160 fb$^{-1}$
(about one and a half year of LHC operation at the high luminosity of
$10^{34}$cm$^{-2}$s$^{-1}$) with approximately the same number of background
events.

The final state $\gamma\gamma+jet$, with high $p_t$ and small rapidity jet
recoiling against the Higgs boson, has not been analysed and we discuss it in
detail in Section 4. The possibility of charged particle tracks reconstruction
in the central detector is an advantage of this signature ($E^{jet}_t>40$ GeV
and $|\eta_{jet}|<2.4$). Final state jet reconstruction allows to determine
more precisely the position of interaction vertex, thus giving the
possibilities to improve two photon invariant mass resolution. This point could
be important for the separation of background. We calculate all the signal
subprocesses and the irreducible background applying realistic kinematical
cuts.The reducible QCD background is estimated as well. 

We calculated cross sections and distributions with the help of CompHEP package
\cite{comphep,IKP}. Exact one-loop matrix elements for the signal QCD
subprocesses were implemented in the CompHEP FORTRAN output, thus we got a
possibility to use CompHEP numerical module for integration over the phase
space.  Then, the vertex $ggH$ in the signal QCD  diagrams, including  quark
loops, was implemented in CompHEP as an effective point-like vertex together
with the $gggH$ vertex to ensure gauge inivariance  of the amplitude. Thus we
tested an accuracy of the effective Lagrangian approximation. 

The CompHEP package includes the code of adaptive  MC integrator VEGAS
\cite{vegas}. For parton distribution functions we used  the parametrizations
CTEQ3m \cite{cteq3}, CTEQ4m and CTEQ4l \cite{cteq4}, MRS-A' \cite{mrs} and
GRV92 (HO) \cite{GRV92}. The corresponding codes are also implemented in
CompHEP. 

The paper is organized as follows. In Section 2 we list the values of physical
parameters and  some physical conventions.

In Section 3 we discuss basic formulas for the $H\to\gamma\gamma$ branching   
and the $ggH$ vertex. To test the code written by using these basic formulas we
compare results obtained by means of CompHEP, PYTHIA 6.1 \cite{pythia} and the
programs HDECAY/HIGLU \cite{HIGLU-HDECAY} for $H\to\gamma\gamma$ and for the
reaction $pp\to H\to\gamma\gamma$.

In Section 4 we analyse the reaction $pp\to\gamma\gamma+jet$.  In Section 4.1
the signal QCD subprocesses $gg\to H+g$, $gq\to H+q$ and $q\bar q\to H+g$ are
discussed. The corresponding formulas for their matrix elements in the limit of
heavy $t$-quark mass  are given, what turns out to be a precise enough
approximation.  Detailed analysis of the main signal and background
subprocesses, $gg\to H+g\to\gamma\gamma+g$ and $gq\to\gamma+\gamma+q$, is
presented in Section 4.2. We introduce different cuts looking for the
possibilities to get a better significance of the Higgs signal in the
$\gamma\gamma+jet$ final state. In Section 4.3 the contributions of all signal
and irreducible background subprocesses are discussed. Then, in Section 4.4, we
estimate the QCD reducible background. In Section 4.5 a general discussion for
the reaction $pp\to\gamma\gamma+jet$ is given. 

In Conclusion the prospects to search for the light Higgs boson at low LHC
luminosity are summarised.

%==================================================================
\section{Input parameters and physical conventions}

In our calculations we used $\sqrt{s}=14$ TeV and $M_Z=91.1884$ GeV,
$\sin^2{\theta_w}=0.2237$. For heavy quark masses we took the values $m_s=0.2$
GeV, $m_c=1.3$ GeV, $m_b=4.3$ GeV and $m_t = 175$ GeV \cite{PDG} \footnote{We
shall not discuss the sensitivity of our results to the value of  $t$-quark
mass. Recent CDF/D0 data gives $m_t=175.6\pm 5.5$ GeV \cite{top}. Rather small
uncertainty of $m_t$ does not change significantly cross sections of the
processes under consideration.}.

Most of the hard subprocesses that we are discussing involve QCD vertices. For
the strong coupling constant we are using the normalization 

\centerline{$\alpha_s(M_Z)=0.118$,} 

\noindent 
where $\alpha_s$ runs in correspondence with the 2nd order QCD (NLO) formulas
with  flavour matching (see, e.g., \cite{PDG}, p.77). This normalization
assumes that $\Lambda^{(5)}_{QCD}=226$ MeV. For instance,
$\alpha_s(120\mbox{\small GeV})=0.1133$. 

In electroweak vertices we used $\alpha(M_Z)=1/128.9$ \cite{PDG} calculating
electroweak couplings and the Fermi constant by means of the corresponding formulas
with $\alpha(M_Z)$, $M_Z$ and $\sin^2{\theta_w}$ as independent parameters.
However, it is more accurate to use $\alpha(m_e)=1/137.036$ in each vertex that
includes an on-shell photon because of the absence of photon wave function  
renormalization in this case. Thus we introduce the correction factor 
$(128.9/137.036)^2$ for each process with two photons in the final state.

%===========================================================================
\section{Higgs decays and the effective Lagrangian}

It is well known that in the case when $M_H<140$ GeV the Higgs boson  decays
dominantly into a $b\bar b$ pair. However, the QCD corrections to this decay
partial width are large. So, for reactions where the $H\to\gamma\gamma$ decay
occurs it is important to take into account carefully these corrections because
they increase the two-photon branching approximately by a factor of 2. In this
section we also discuss all other decay channels related to the 
$H\to\gamma\gamma$ branching and the effective $\gamma\gamma H$ Lagrangian
which can be used for calculation of the partial width
$\Gamma_{H\to\gamma\gamma}$. Similar effective $ggH$ Lagrangian can be used in
calculation of the partial width of Higgs decay into a gluon pair and the cross
sections of signal QCD processes. 

In order to take into account the QCD corrections the NNLO formulas improved by
the renormalization group method \cite{MQrunning} were proposed for the running
quark masses. We implemented in CompHEP the corresponding NLO formulas for the
Higgs decay into quarks. This approximation gives results for partial widths
higher than the NNLO formulas by $\sim 10$\% and the corresponding correction
factor has been introduced in our computer code.

Electroweak corrections to the Higgs decays into quarks and leptons are small
\cite{EW-Hdec} and we neglect them.

Analytical formulas for the partial widths $\Gamma_{H\to\gamma\gamma}$ and 
$\Gamma_{H\to gg}$ in the leading one-loop approximation are known for many 
years \cite{effLagr}. They can be obtained with the help of the  following
effective Lagrangians:
$$
{\cal L}^{eff}_{\gamma\gamma H}=
        \frac{\lambda_{\gamma\gamma H}}{2} F_{\mu\nu} F^{\mu\nu} H\;,
$$
\begin{equation}
{\cal L}^{eff}_{ggH}=K \frac{\lambda_{ggH}}{2} 
    G^a_{\mu\nu} G^{a\mu\nu} H\;. 
                                          \label{eq:ggh-lagr}
\end{equation}
Here $F_{\mu\nu}$ and $G^a_{\mu\nu}$ are photon and gluon field strengths,
correspondingly, and $K$ is a factor accounting for high order QCD corrections.
Analytical formulas for the effective coupling constants 
$\lambda_{\gamma\gamma H}$ and $\lambda_{ggH}$ in the leading order
approximation can be found in Refs.~\cite{effLagr}.  The corresponding partial
widths are
$$ \Gamma_{H\to\gamma\gamma} = 
    \frac{\lambda_{\gamma\gamma H}^2}{16\pi} \cdot M_H^3\;,
  \qquad
\Gamma_{H\to gg} = K^2 \cdot
    \frac{\lambda_{ggH}^2}{2\pi} \cdot M_H^3\;.
$$  

Total QCD radiative corrections to the $H\to gg$ partial width can be collected
in a factor with value $\sim 1.7$ \cite{Dawson91,DSZ,SDGZ,Chet}. Note that the
QCD radiative corrections to the partial width of $H\to\gamma\gamma$ are very
small (less than 1\%, see, e.g. \cite{SDGZ} and references therein).

Electroweak corrections to the partial widths of $H\to\gamma\gamma$ and $H\to
gg$ are small and can be neglected \cite{EW-aa-gg}.

In the limit $m_t\to\infty$ and without contributions from the  light quark
loops the $ggH$ effective coupling constant has a very simple form
\begin{equation}
 \lambda_{ggH} \Bigl|_{m_t\to\infty} \;=\; \frac{\alpha_s g_w}{12\pi M_W}\;,
                    \label{eq:lambdaeff}
\end{equation}
where the electroweak constant is
$\alpha_w=g_w^2/4\pi=\alpha(M_Z)/\sin^2{\theta_w}$.  As we shall see the
effective Lagrangian (\ref{eq:ggh-lagr}) with the coupling constant
(\ref{eq:lambdaeff}) can be used in calculations of the signal subprocesses
with acceptable accuracy.  

We remark that for a precise calculation of the $H\to\gamma\gamma$
branching in the Higgs mass range considered one has to account also for the
partial width  $\Gamma_{H\to WW^*}$, where $W^*$ means the off-shell $W$-boson.
At the mass values close to $M_H=140$ GeV
this branching and the branching to a $b\bar b$ pair are of the same order. The
corresponding analytical formulas can be found in \cite{WWstar}.

All formulas for the partial widths mentioned above and the higher order QCD
correction factors were implemented in the CompHEP codes. We checked the
numbers  obtained for the branching $H\to\gamma\gamma$ by means of CompHEP,
PYTHIA 6.1 and the program HDECAY \cite{HIGLU-HDECAY}:

\begin{center}
\begin{tabular}{cccc}
&\multicolumn{3}{c}{$Br(H\to\gamma\gamma)\cdot 10^{-3}$}\\ 
\hline
 $M_H$  GeV   &  CompHEP   & HDECAY  &PYTHIA \\ 
\hline
  100         &  1.999      & 1.961       & 2.065  \\
  120         &  2.723      & 2.707       & 2.857  \\
  140         &  2.148      & 2.151       & 2.327  \\
\hline
\end{tabular}
\end{center}

One can see a good agreement of our numbers with the  results of the HDECAY
program (where all known formulas for higher order corrections are used).  At
the same time one can see that the PYTHIA results are 5-8\% higher.

To test our code written  for the effective coupling constant  $\lambda_{ggH}$
(in leading order) we compared also the numbers obtained for the $pp\to
H\to\gamma\gamma$ cross section by means of CompHEP, PYTHIA 6.1 and
HIGLU/HDECAY programs:  

\begin{center}
\begin{tabular}{llccc}
& &
\multicolumn{3}{c}{$\sigma^{tot}(gg\to H\to\gamma\gamma)$, fb} \\
\hline
package  & PDF set   &   $M_H=100$ GeV  & 120 GeV & 140 GeV  \\
\hline
 CompHEP & CTEQ3m              & 37.18  & 41.03  & 26.85  \\
         & GRV92(HO)           & 38.08  & 41.11  & 26.45  \\
                               & & & \\
HIGLU/HDECAY& GRV92(HO)        & 39.59  & 40.78  & 25.17   \\
                               & & & \\
PYTHIA 6.1 & CTEQ3m            & 38.96  & 40.97  & 26.65   \\
\hline
\end{tabular}
\end{center}

Here we used $Q^2=M_H^2$ for the parton factorization scale. The strong
coupling  was taken fixed at the value of Higgs mass. The hard subprocess
$gg\to H$ was calculated in leading order  $\alpha_s^2$. Values of the 
$H\to\gamma\gamma$ branching used in these programs are given above. One can
see that all three programs are in agreement with each other within 5\%.

%===========================================================================
\section{\boldmath Higgs boson signal in the reaction $pp\to\gamma\gamma+jet$}

In this section we study systematically processes with the final state
$\gamma\gamma+jet$. 

First of all let us note that the signal subprocesses contributing to this
reaction are QCD subprocesses, and it means that the cross sections should
depend strongly on QCD parameters. There are three main sources of this
dependence. The first one is defined by the evolution of the parton densities.
Then, the hard subprocess cross sections depend on the  running $\alpha_s$. 
Finally, the $H\to\gamma\gamma$ branching depends on the Higgs total width,
where the QCD corrections are large (see a short discussion  in Section 3 and
the corresponding references). In leading order all these three sources can be
factorized. Moreover, due to a very small value of $\Gamma^{tot}_H$ in the
Higgs mass range of 100-140 GeV the decay width of $H\to\gamma\gamma$ can be
factorized, and the fixed value of strong coupling $\alpha_s(M_H)$ can be used
for evaluation of the $H\to\gamma\gamma$ branching. However, it is well-known
that for the reaction $pp\to H\to\gamma\gamma$ the dependence on the QCD
normalization scale $\mu$ (used in calculations of the QCD corrections to hard
subprocesses) and on the parton factorization scale $Q$ is strong enough and
next-to-leading analysis is necessary (see, e.g. \cite{SDGZ} and references
therein). For the reaction $pp\to H\to\gamma\gamma$ after including the NLO
corrections, this theoretical uncertainty decreases considerably, showing only
a $\sim 15$\% remaining $(\mu,Q)$ sensitivity. Surely one can expect a similar
effect also for the Higgs production at high $p_t$. The self-consistent
analysis requires the NLO corrections to hard subprocesses which are not known
yet. Thus, today we cannot analyse the reaction $pp\to\gamma\gamma+jet$ in the
complete NLO approximation. Therefore, in  this paper some kind of combined
accounting for the QCD effects is used: NLO PDF evolution and LO approximation
for hard subprocesses (but with NLO running $\alpha_s$). 

%----------------------------------------------------
\subsection{Signal QCD subprocesses}

    There are three hard QCD subprocesses with a signal from the
Higgs boson contributing to the reaction $pp\to\gamma\gamma+jet$:
\begin{equation}
                    gg\to H+g\;,     \qquad
                    gq\to H+q\;,     \qquad
               q\bar q\to H+g.     \label{eq:QCDsignal} 
\end{equation}

In Fig.~\ref{fig:fd_s_QCD} Feynman diagrams contributing in leading  order 
$\alpha_s^3$ are shown. Here the vertices $ggH$ and $gggH$ are substituted
instead of the corresponding quark loops in exact calculations and
represent the effective point-like vertices of the Lagrangian
(\ref{eq:ggh-lagr},\ref{eq:lambdaeff}) in the approximate calculations.

The corresponding matrix elements were calculated analytically in \cite{j1}.
In the limit $m_t\to\infty$ the differential cross sections are 
\begin{equation}
 \frac{d\sigma_{gg}}{d{\hat t}}  \Bigl|_{m_t\to\infty}  \;=\;
  \frac{1}{16\pi s^2} \cdot
  \frac{\alpha_w \alpha_s^3}{24} 
  \left[ \frac{M_H^8+{\hat s}^4+{\hat t}^4+{\hat u}^4}
           {{\hat s}{\hat t}{\hat u} M_W^2} \right] \;,
                                        \label{eq:SQMEsignal}
\end{equation}
$$ \frac{d\sigma_{gq}}{d{\hat t}}\Bigl|_{m_t\to\infty}  \;=\; 
   \frac{1}{16\pi s^2}\cdot 
   \frac{\alpha_w \alpha_s^3}{54} \cdot 
   \frac{{\hat s}^2 + {\hat u}^2}{(-{\hat t}) M_W^2}\;,
 \qquad
   \frac{d\sigma_{q\bar q}}{d{\hat t}} \Bigl|_{m_t\to\infty} \;=\; 
   \frac{1}{16\pi s^2} \cdot
   \frac{4 \alpha_w \alpha_s^3}{81} \cdot 
   \frac{{\hat t}^2 + {\hat u}^2}{{\hat s} M_W^2}\;.
$$
Here $\hat s$, $\hat t$ and $\hat u$ are the standard Mandelstam variables  for
partons in their c.m. system. These formulas can be derived from the effective
Lagrangian (\ref{eq:ggh-lagr},\ref{eq:lambdaeff}).

We found that the approximation (\ref{eq:SQMEsignal}) works well enough.  For
the channel $gg$ this approximation gives numbers slightly smaller than the
exact matrix elements -- the difference is less than 7\% at the transverse
momentum cut $p_t>40$ GeV and $|\eta_{jet}|<2.4$. For the channel $gq$ this
approximation overestimates the exact result by $\sim 13$\% with the same cuts.
For the channel $q\bar q$  the approximation (\ref{eq:SQMEsignal})  does not
work at all, however,  the contribution of this channel to the process
$pp\to\gamma\gamma+jet$ is much smaller than the contribution of $gg$ and $gq$
channels and can be neglected. Thus, the precision of the effective Lagrangian
approximation in calculation of the QCD signal processes (\ref{eq:QCDsignal})
is better than 7\%. This fact allows in principle to create a fast generator of signal
events for the reaction $pp\to\gamma\gamma+jet$. Detailed analysis of the
approximation (\ref{eq:SQMEsignal}) will be published elsewhere
\cite{ggH-approx}.

We checked that the PYTHIA result is also in good agreement with exact
calculations, giving cross section  higher by $\sim 6$\%.

%----------------------------------------------------
\subsection{\boldmath $pp\to H+g\to\gamma\gamma +g$ {\it versus} 
$pp\to\gamma+\gamma+q$}

In this section we consider the contribution of the signal subprocess  $gg\to
H+g\to\gamma\gamma+g$ (four diagrams in Fig.1a). We compare the corresponding
cross section and distributions with those of the background subprocess
$gq\to\gamma+\gamma+q$, $q=u,d$. Tree level Feynman diagrams
contributing to this background are shown in Fig.~\ref{fig:fd_Birr}a. Let us
stress that the two subprocesses considered give the main contributions to the
signal and background. In this section we look for the possibilities to get a
better significance of the Higgs signal varying the cuts on different
variables.

In the following we consider realistic cuts \cite{cmsTP,atlasTDR} for the final
state $\gamma\gamma+jet$  (which we shall refer as to the
{\bf (C)} set of cuts):\footnote{$\Delta R=\sqrt{\Delta\eta^2+\Delta\phi^2}$ is
a separation between two particles in the {\it azimuth angle -- rapidity}
plane.}

\begin{itemize}
\item[\bf (C1)] two photons are required with $p_t^\gamma>40$ GeV 
    and $|\eta_\gamma|<2.5$ for each photon;
\item[\bf (C2)]  photons are isolated from each other by 
  $\Delta R(\gamma_1,\gamma_2) > 0.3$;
\item[\bf (C3)] jet has high transverse energy $E_t^{jet}>40$ GeV and is 
  centrally produced, $|\eta_{jet}|<2.4$;
\item[\bf (C4)] jet is isolated from the photons by
  $\Delta R(jet,\gamma_{1})>0.3$ and $\Delta R(jet,\gamma_{2})>0.3$.
\end{itemize}

In the following we shall vary the value of the kinematical cut on one of the
variables starting from this basic set and leave a more tuned multivariable
analysis for the future.

It is clear that the signal significance depends strongly on the two-photon
invariant mass resolution. This key instrumental characteristic at a low luminosity
is expected to be $\sim 1.3$ GeV for the CMS electromagnetic calorimeter
\cite{cmsTP} and $\sim 3.1$ GeV for the  ATLAS calorimeter \cite{atlasTDR} 
($M_H=100$ GeV). These numbers are for the case of $\sim 80$\%
signal events reconstruction. Hereafter we shall integrate the background  over
$\gamma\gamma$  invariant mass  within the range $M_H-\Delta
M_{\gamma\gamma}<M_{\gamma\gamma}<M_H+\Delta M_{\gamma\gamma}$ with $\Delta
M_{\gamma\gamma}=1$ GeV. 

Cross sections of the  subprocesses under discussion convoluted with various
parametrizations of the parton distributions for the basic set of kinematical
cuts {\bf (C)} and for $M_H=120$ GeV are

\begin{center}
\begin{tabular}{lccc}
                                             &CTEQ4m&CTEQ3m& MRS-A' \\
 \hline
 {\bf (S)} ,\hspace{0.3cm} fb                & 4.32 & 4.65 & 4.44   \\
 {\bf (B)}{\Large /}2GeV ,\hspace{0.3cm} fb  & 11.89& 11.94 & 11.94  \\ 
\hline
\end{tabular}
\end{center}

Thereinafter we denote by {\bf (S)} the signal process $pp\to
H+g\to\gamma\gamma+g$ and by {\bf (B)} the background process
$pp\to\gamma+\gamma+q$. Here we use $Q^2=M_H^2+2(E_t^{jet})^2$ as the parton
factorization scale.

To estimate the NLO corrections we calculated these cross sections also with
the leading order parametrization CTEQ4l and with strong coupling running in
leading order. The LO results appear to be $\sim 15$\% lower for the signal and
only $\sim 3$\% lower for the background. For the signal about 2/3 of this
difference is due to the running of $\alpha_s$, while for the background
practically all the decrase is related to the PDF evolution.

%\begin{center}
%\begin{tabular}{lcccc}
%& CTEQ4m-$\alpha_s^{NLO}$ & CTEQ4l-$\alpha_s^{LO}$ 
%& CTEQ4m-$\alpha_s^{LO}$  & CTEQ4l-$\alpha_s^{NLO}$ \\
% \hline
% {\bf (S)} ,\hspace{0.3cm} fb                & 4.32 & 3.74 & 3.95 & 4.07  \\
% {\bf (B)}{\Large /}2GeV ,\hspace{0.3cm} fb  & 11.89& 11.70& 11.96& 11.73 \\ 
%\hline
%\end{tabular}
%\end{center}

The parton c.m. energy distributions presented in Fig.~\ref{fig:aaj-shat} show
that almost all events have a small parton c.m. energies $\sqrt{\hat s}\le 300$
GeV.  So the phase space region for the process under discussion is characterized by
the relations $\sqrt{\hat s}\sim M_H\sim m_t$. It means that we have events
with very small values of the Bjorken variable $x\sim 3\cdot 10^{-4}$. For
these small $x$ values the recent DIS data from HERA \cite{HERAdata} show some
discrepancy with the existing parametrizations, especially for the gluon
distribution. This data, as well as the recent DIS data from the NMC and  E665
experiments and from precision measurements of the inclusive jet production at
Tevatron, were used to improve the parton distributions by the CTEQ
Collaboration \cite{cteq4}. We have found that its latest parametrization
CTEQ4m gives results lower than the previous set CTEQ3m by $\sim 7.5$\% for the
signal and only by $\sim 0.5$\% for the background subprocesses (see the Table
above). We use the CTEQ4m parametrization ($\alpha_s(M_Z)=0.116,\;
\Lambda^{(5)}_{QCD}=202$ MeV) in our calculations. Furthemore, we calculated
the cross sections with different choices of the $Q^2$ parameter which was used
for both the parton factorization scale and the normalization for running
$\alpha_s$ (numbers in the Table below are for $M_H=120$ GeV): 

\begin{center}
\begin{tabular}{lccc}
                 & $Q^2=M_H^2+2(E_t^{jet})^2$,  & $M_H^2,$ & $(E_t^{jet})^2$ \\
 \hline
 {\bf (S)} ,\hspace{0.3cm} fb               & 4.32 & 4.99 & 5.73   \\
 {\bf (B)}{\Large /}2GeV ,\hspace{0.3cm} fb &11.89 & 12.18& 12.68  \\ 
\hline
\end{tabular}
\end{center}
One can see a much stronger dependence on the choice of QCD parameter $Q^2$
for the signal  cross sections than for the background ones. Of course, it is
caused by higher QCD order ($\alpha_s^3$) of the signal subprocesses, while the
background is only of order of $\alpha_s$.  In the following we use 
$Q^2=M_H^2+2(E_t^{jet})^2$ which is intuitively more relevant  to  physics
discussed here. 

Let us look at the rapidity interval between the jet and the Higgs boson due to
which the minijet  corrections \cite{DucSch94} could give some enhancemenet
factor if one considers the semi-inclusive process $pp\to H+jet+X$. In
Fig.~\ref{fig:gapeta} the distribution in the rapidity interval $(y_{jet}-y_H)$
is presented. One can see that this interval is less than $\pm 3$. So 
the corresponding minijet corrections should be small unlike in the case
of SSC discussed in \cite{DucSch94}. 

Then, in Fig.~\ref{fig:aaj-etag} one can see that varying the cuts on photon
rapidities around set {\bf (C)} values does not help to improve signal
significance. However, a change of the jet rapidity acceptance does show some
prospects (Fig.~\ref{fig:aaj-etaj}). In Table~\ref{tab:aaj-jcuts}(a)
\footnote{Here the {\it effective} significance $\sigma_S/\sqrt{\sigma_B}$ is
presented rather than  usually used $N_S/\sqrt{N_B}$. To get the latter one
should multiply the effective significance by the square root of the integrated
luminosity.} one can see that the significance increases when the jets with
larger rapidities are detected. For example, we get a $\sim 14$\% increase of
significance  if the $|\eta^j|<2.4$ cut is replaced by $|\eta^j|<4$ with
involving the very forward hadron calorimeter.

The distribution in the jet transverse energy for the two processes discussed
are shown in Fig.~\ref{fig:aaj-etj} for two values of the photon  transverse
momentum cut $p_t^\gamma >20$ and 40 GeV. One can see that the application of a
stronger cut improves the $S/B$ ratio. This ratio increases also for a stronger
$E_t^{jet}$ cut (see Fig.~\ref{fig:aaj-ptg}).\footnote{Some suppression in the
interval $(M_H-\bar E_t^{jet})/2<p_t^\gamma<(M_H+\bar E_t^{jet})/2$ is
connected with the cut applied on the jet transverse energy $E_t^{jet}>\bar
E_t^{jet}$.} Let us see, however, what significance do we have. We have found
that the variation of $p_t^\gamma$ around set {\bf (C)} values does not help to
improve it. However, the application of a weaker cut on the analogous jet
variable $E_t^{jet}$ does improve the signal significance. For example, a
replacement of the cut  $E^{jet}_t>40$ GeV by $E^{jet}_t>30$ GeV increases the
significance by $\sim 8$\% but leads to a small decrease of the ratio $S/B$,
see Table~\ref{tab:aaj-jcuts}(b).

We analysed also a variety of other cuts, in particular, the cuts on 
transverse momenta of the photons, their relative angle in the two-photon rest
frame and their separation cut. We treated also the cut on photon pair
transverse momentum $p_t^{\gamma_1+\gamma_2}$, and different cuts related to
the planarity features of events. Unfortunately these variables did not show
any possibilities to get a better signal significance.

%--------------------------------------------------------------------------
\subsection{Other signal and (irreducible) background processes}

A number of other subprocesses contribute to the reaction 
$pp\to\gamma\gamma+jet$, both to the signal and background. We have calculated
all of them within the framework described above. Results are collected in
Table~\ref{tab:aaj-all}.

Signal subprocesses can be subdivided in two groups. In the first group there
are QCD subprocesses (\ref{eq:QCDsignal}) discussed in Section 4.1. Numerically
the $gq$ channels give about 12\% of the main $gg$ signal  contribution, while
the $q\bar q$ channels can be neglected. 

\begin{figure}[hb]
\begin{center}
\unitlength=1cm
\vspace*{7.0cm}
\begin{picture}(16,10)
\put(-1,-0.5){\epsfxsize=16cm \leavevmode \epsfbox{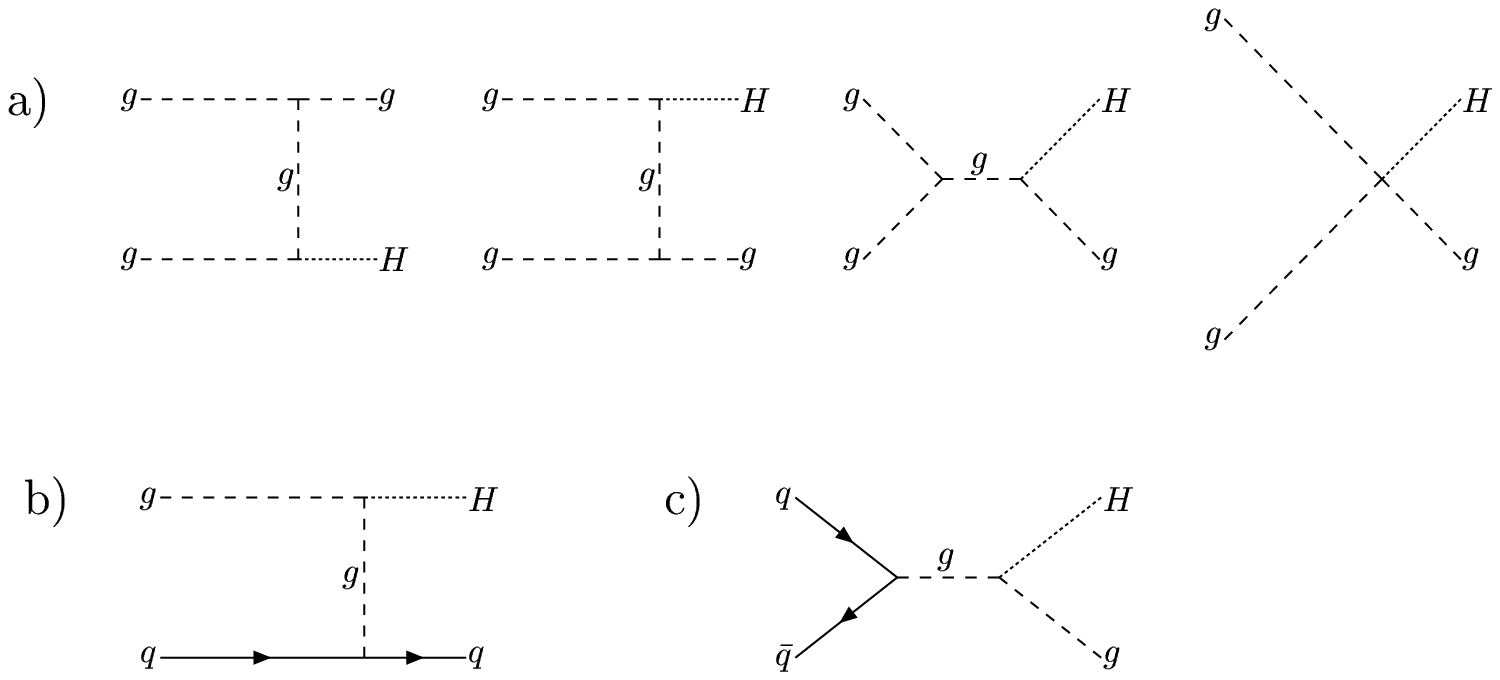}}
\end{picture}
\end{center}
\vspace*{-14cm}
\caption{Feynman diagrams for the QCD signal subprocesses:
a) $gg\to H+g$ , b) $gq\to H+q$ and c) $q\bar q\to H+g$.
In SM the $ggH$ and $gggH$ vertices include quark loops. In the
effective Lagrangian approximation 
{\protect (\ref{eq:ggh-lagr},\ref{eq:lambdaeff})} 
these vertices are point-like with the coupling constant $\lambda_{ggH}$.
\label{fig:fd_s_QCD} }
%\end{figure}

%\begin{figure}[hb]
\begin{center}
\unitlength=1cm
\vspace*{4cm}
\begin{picture}(16,16)
\put(-1,-0){\epsfxsize=16cm \leavevmode \epsfbox{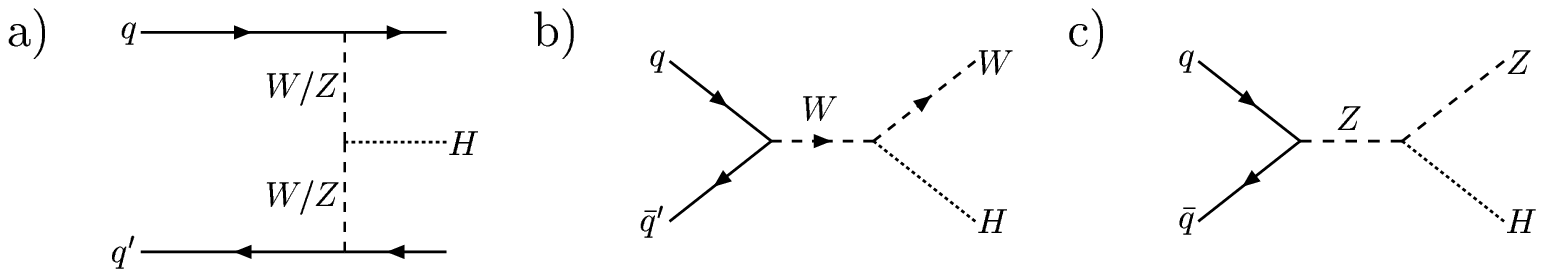}}
\end{picture}
\end{center}
\vspace*{-17cm}
\caption{Feynman diagrams for the signal subprocesses with a) $WW$ and $ZW$
fusion mechanisms of the Higgs boson production, b) associated $W$-boson
and c) associated $Z$-boson production. 
\label{fig:fd_s_WZ} }
%\end{figure}

%\clearpage
%\begin{figure}[hb]
%\unitlength=1cm
\vspace*{20cm}
%\begin{center}
\begin{picture}(16,16)
\put(-1,0){\epsfxsize=16cm \leavevmode \epsfbox{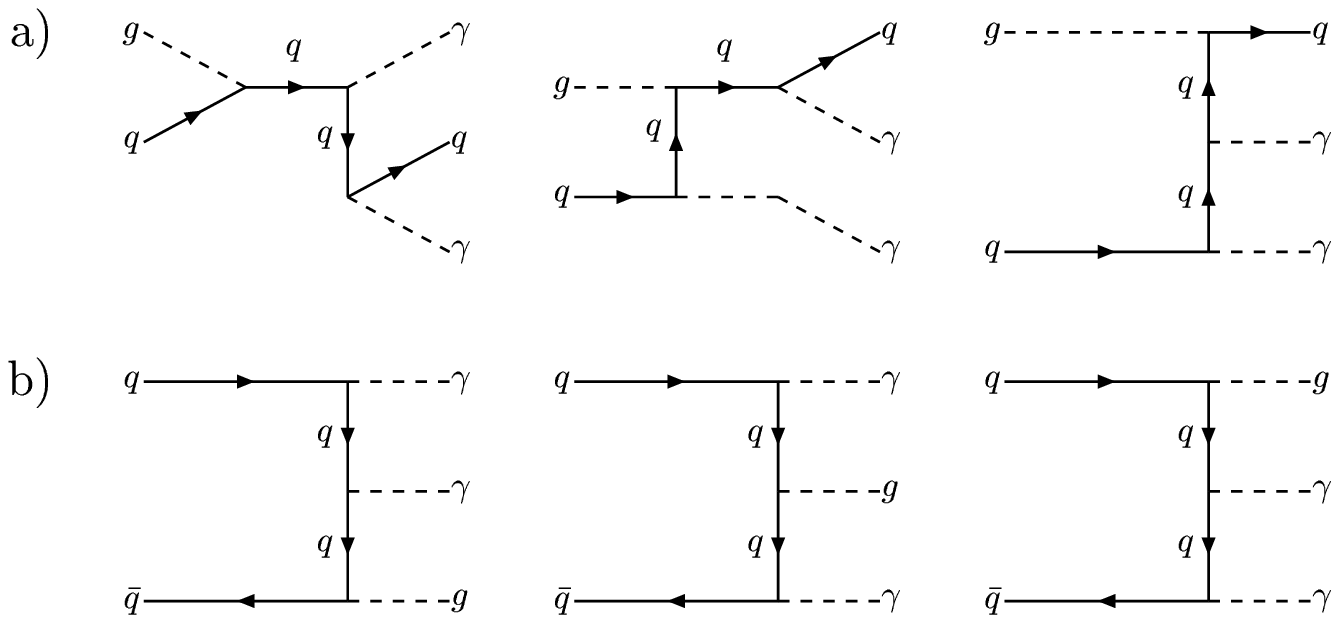}}
\end{picture}
%\end{center}
\vspace*{-14cm}
\caption{Feynman diagrams for the subprocesses
a) $gq\to\gamma+\gamma+q$ and b) $q\bar q\to\gamma+\gamma+g$
contributing to the background.
\label{fig:fd_Birr} }
\end{figure}

\begin{figure}[hb]
\begin{center}
\unitlength=1cm
\begin{picture}(16,16)
\put(0,0){\epsfxsize=16cm \leavevmode \epsfbox{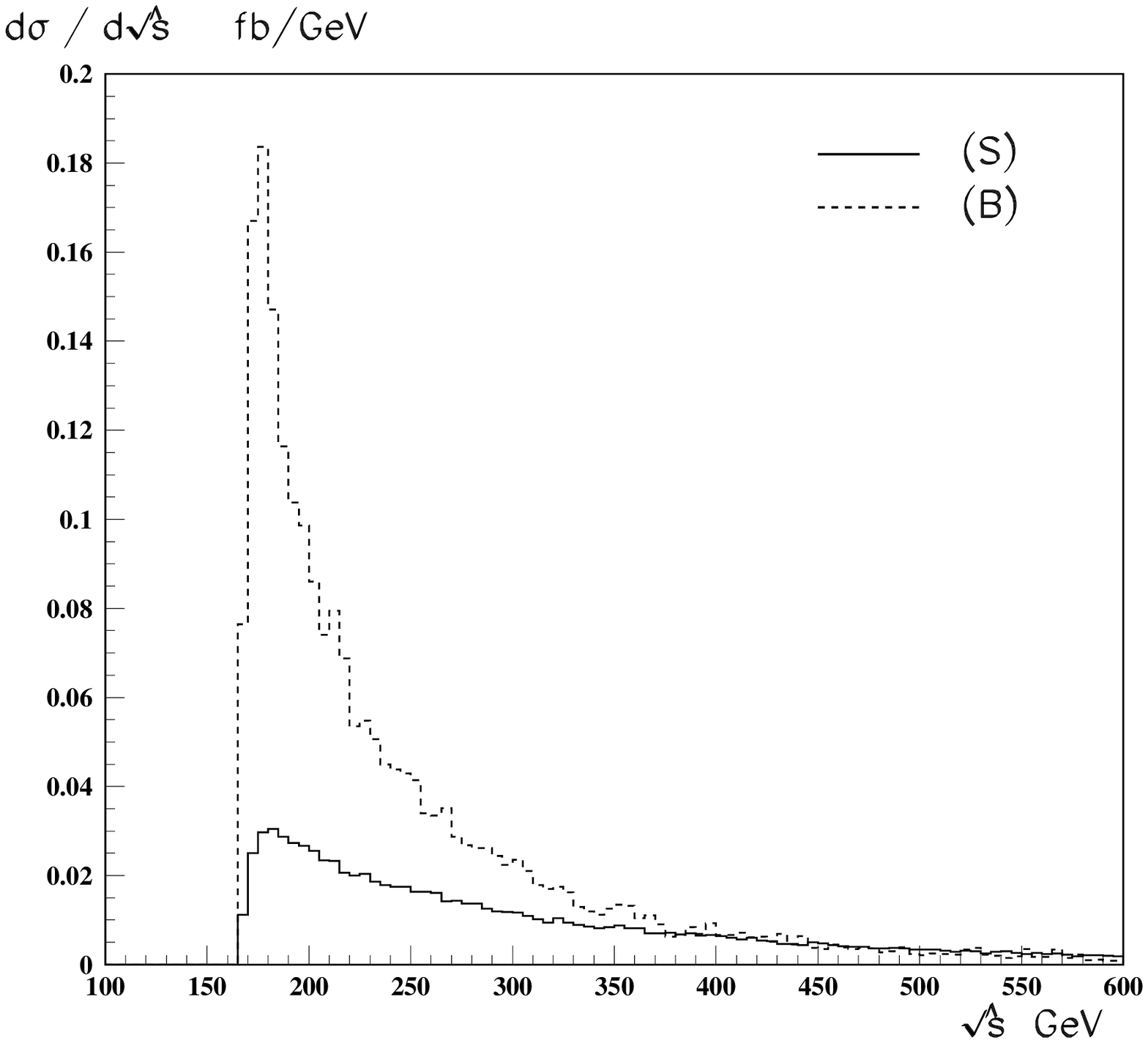}}
\end{picture}
\end{center}
\caption{Distributions in the parton c.m. energy ${\protect \sqrt{\hat s}}$
for the main signal (S) and background (B) processes 
$pp\to H+g\to\gamma\gamma +g$ and $pp\to\gamma+\gamma+q$. 
Energy ${\protect\sqrt{s}}=14$ TeV and the Higgs boson mass $M_H=120$ GeV. 
CTEQ4m PDF set and $Q^2=M_H^2+2(E^{jet}_t)^2$ are used. 
The basic set of kinematical cuts {\bf (C)} is imposed.
The $\gamma\gamma$ invariant mass for the background is integrated over the 
range $M_H-\Delta M_{\gamma\gamma} < M_{\gamma\gamma} < M_H +
\Delta M_{\gamma\gamma}$ with $\Delta M_{\gamma\gamma}=1$ GeV. 
\label{fig:aaj-shat}}
\end{figure}

\begin{figure}[hb]
\begin{center}
\unitlength=1cm
\begin{picture}(16,16)
\put(0,0){\epsfxsize=16cm \leavevmode \epsfbox{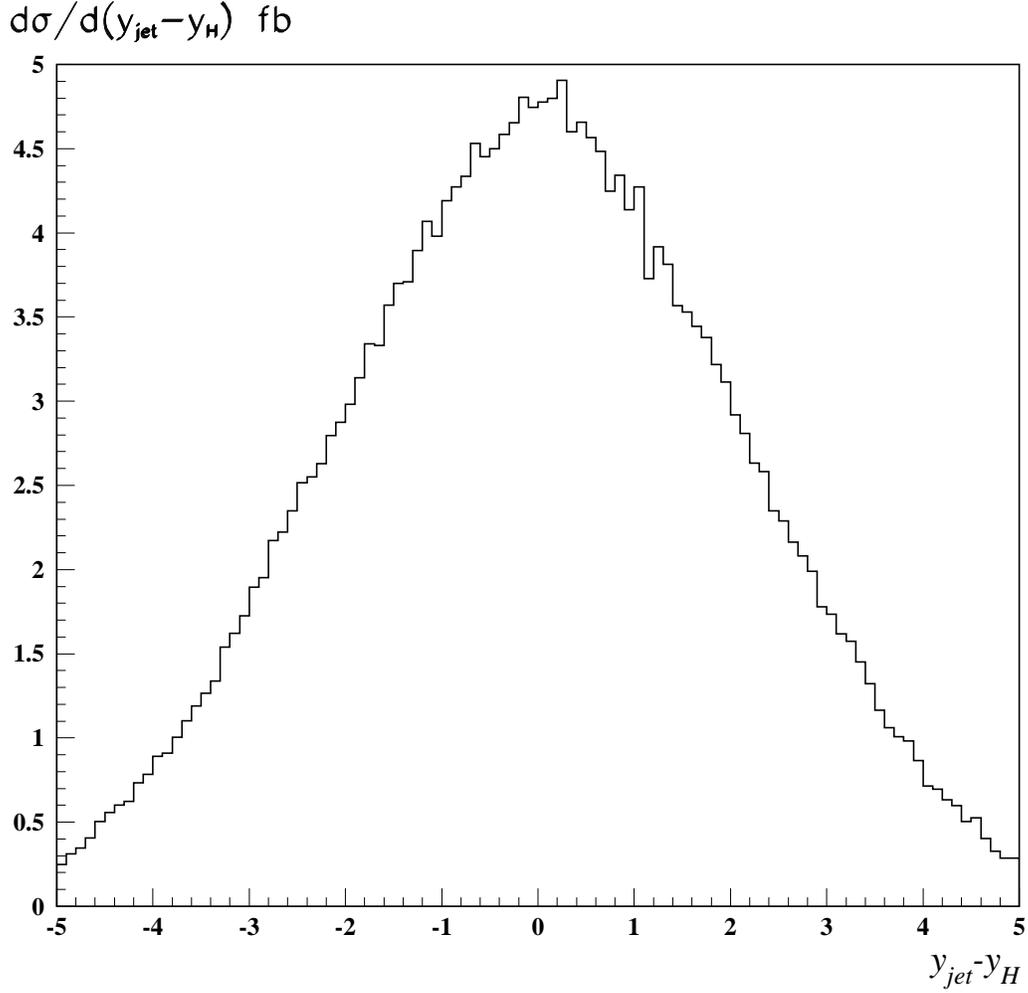}}
\end{picture}
\end{center}
\caption{Distribution in rapidity interval $y_{jet}-y_H$
for the main signal (S) process $pp\to H+g\to\gamma\gamma +g$. 
Energy ${\protect \sqrt{s}}=14$ TeV and the Higgs boson mass $M_H=120$ GeV. 
CTEQ4m PDF set and $Q^2=M_H^2+2(E^{jet}_t)^2$ are used.
The basic set of kinematical cuts {\bf (C)} is imposed.
\label{fig:gapeta}}
\end{figure}

\begin{figure}[hb]
\begin{center}
\unitlength=1cm
\begin{picture}(16,16)
\put(0,0){\epsfxsize=16cm \leavevmode \epsfbox{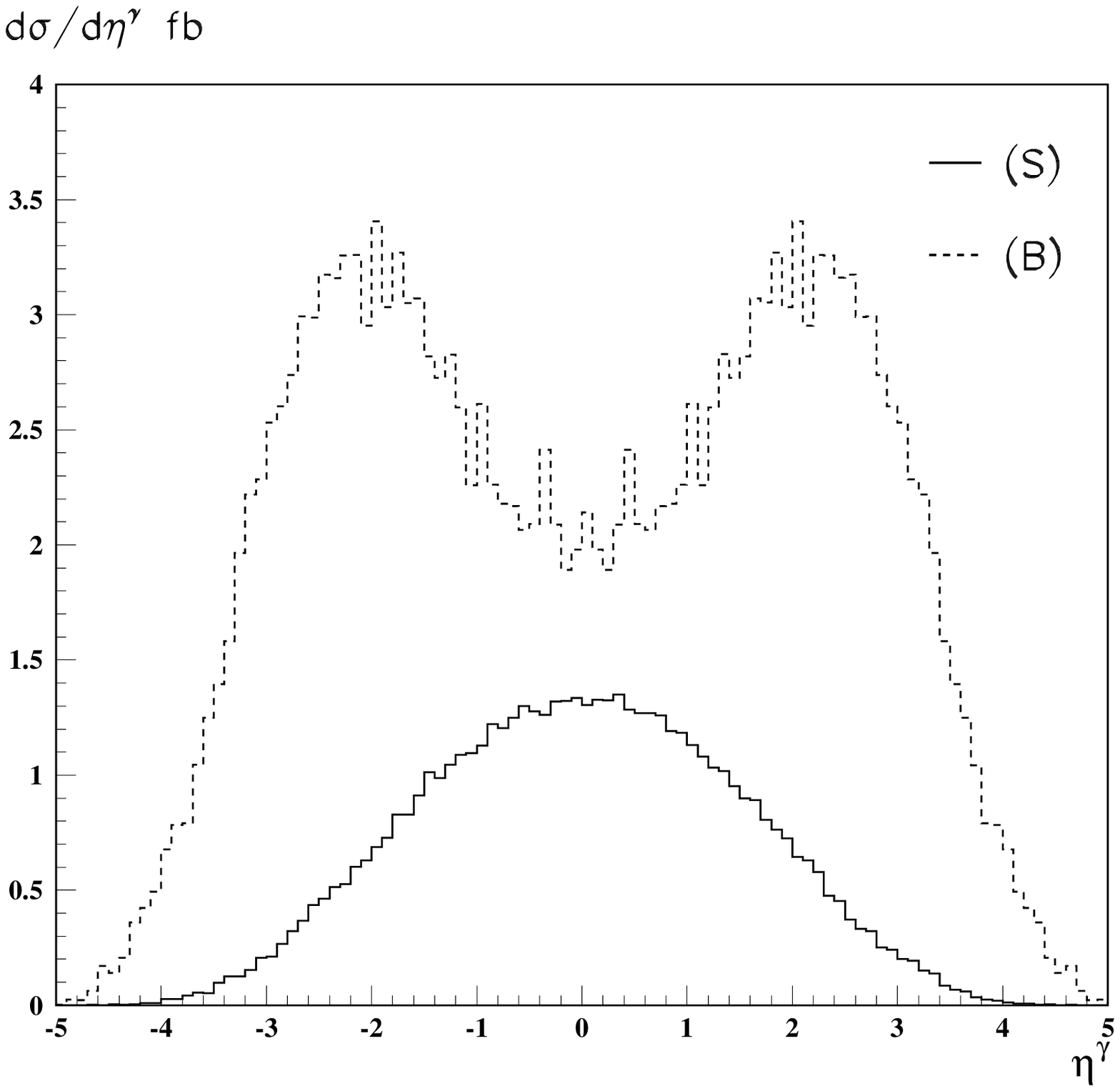}}
\end{picture}
\end{center}
\caption{Distributions in photon rapidity 
for the main signal (S) and background (B) processes 
$pp\to H+g\to\gamma\gamma +g$ and $pp\to\gamma+\gamma+q$. 
Energy ${\protect \sqrt{s}}=14$ TeV and the Higgs boson mass $M_H=120$ GeV. 
CTEQ4m PDF set and $Q^2=M_H^2+2(E^{jet}_t)^2$ are used.
The basic set of kinematical cuts {\bf (C)} is imposed on other variables.
The $\gamma\gamma$ invariant mass for the background is integrated over the 
range $M_H-\Delta M_{\gamma\gamma} < M_{\gamma\gamma} < M_H +
\Delta M_{\gamma\gamma}$ with $\Delta M_{\gamma\gamma}=1$ GeV. 
\label{fig:aaj-etag}}
\end{figure}

\begin{figure}[hb]
\begin{center}
\unitlength=1cm
\begin{picture}(16,16)
\put(0,0){\epsfxsize=16cm \leavevmode \epsfbox{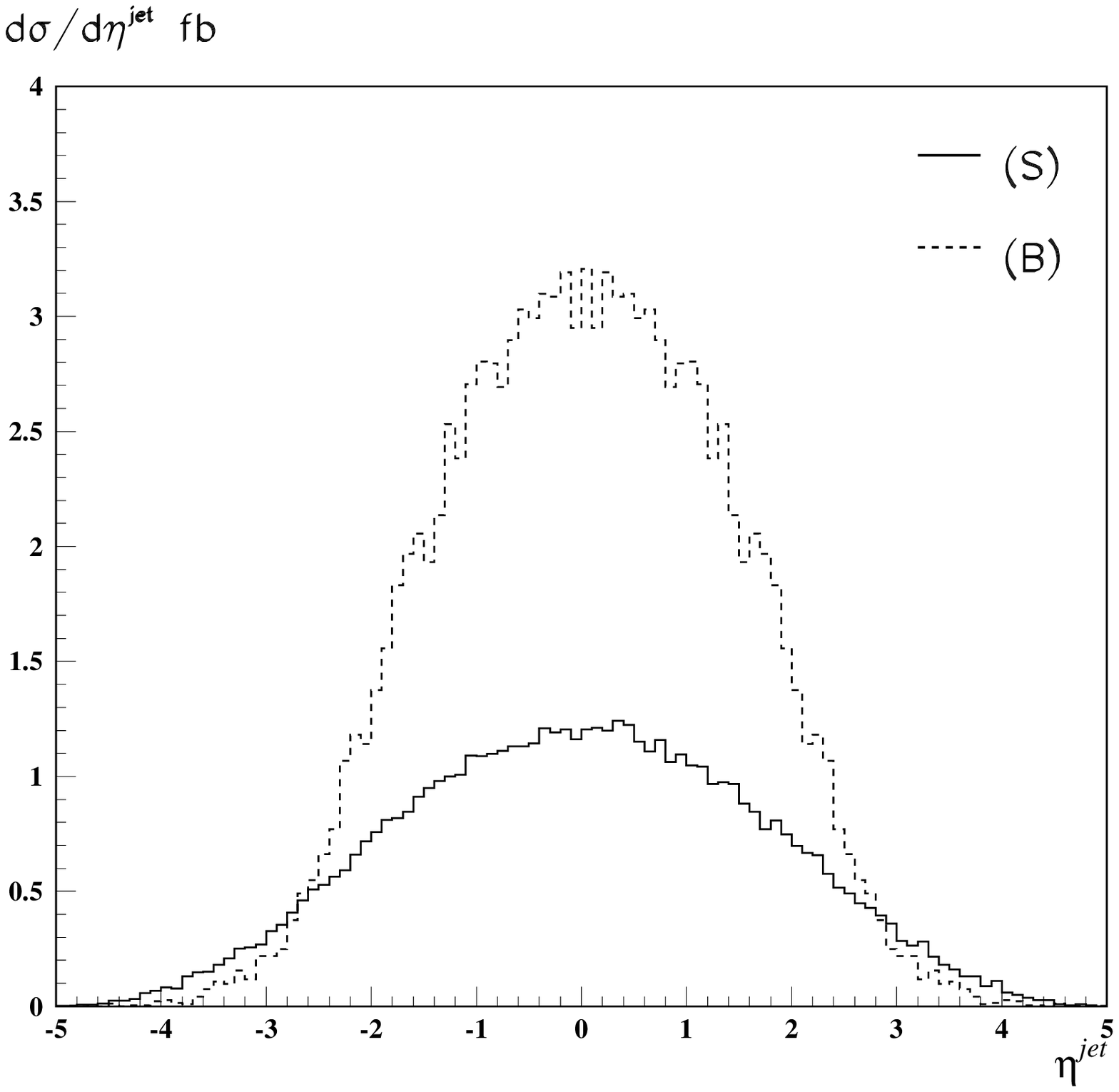}}
\end{picture}
\end{center}
\caption{Distributions in jet rapidity 
for the main signal (S) and background (B) processes 
$pp\to H+g\to\gamma\gamma +g$ and $pp\to\gamma+\gamma+q$. 
Energy ${\protect \sqrt{s}}=14$ TeV and the Higgs boson mass  $M_H=120$ GeV. 
CTEQ4m PDF set and $Q^2=M_H^2+2(E^{jet}_t)^2$ are used.
The basic set of kinematical cuts {\bf (C)} was imposed on other variables.
The $\gamma\gamma$ invariant mass for the background is integrated over the 
range $M_H-\Delta M_{\gamma\gamma} < M_{\gamma\gamma} < M_H +
\Delta M_{\gamma\gamma}$ with $\Delta M_{\gamma\gamma}=1$ GeV. 
\label{fig:aaj-etaj}}
\end{figure}

\begin{figure}[hb]
\begin{center}
\unitlength=1cm
\begin{picture}(16,16)
\put(0,0){\epsfxsize=16cm \leavevmode \epsfbox{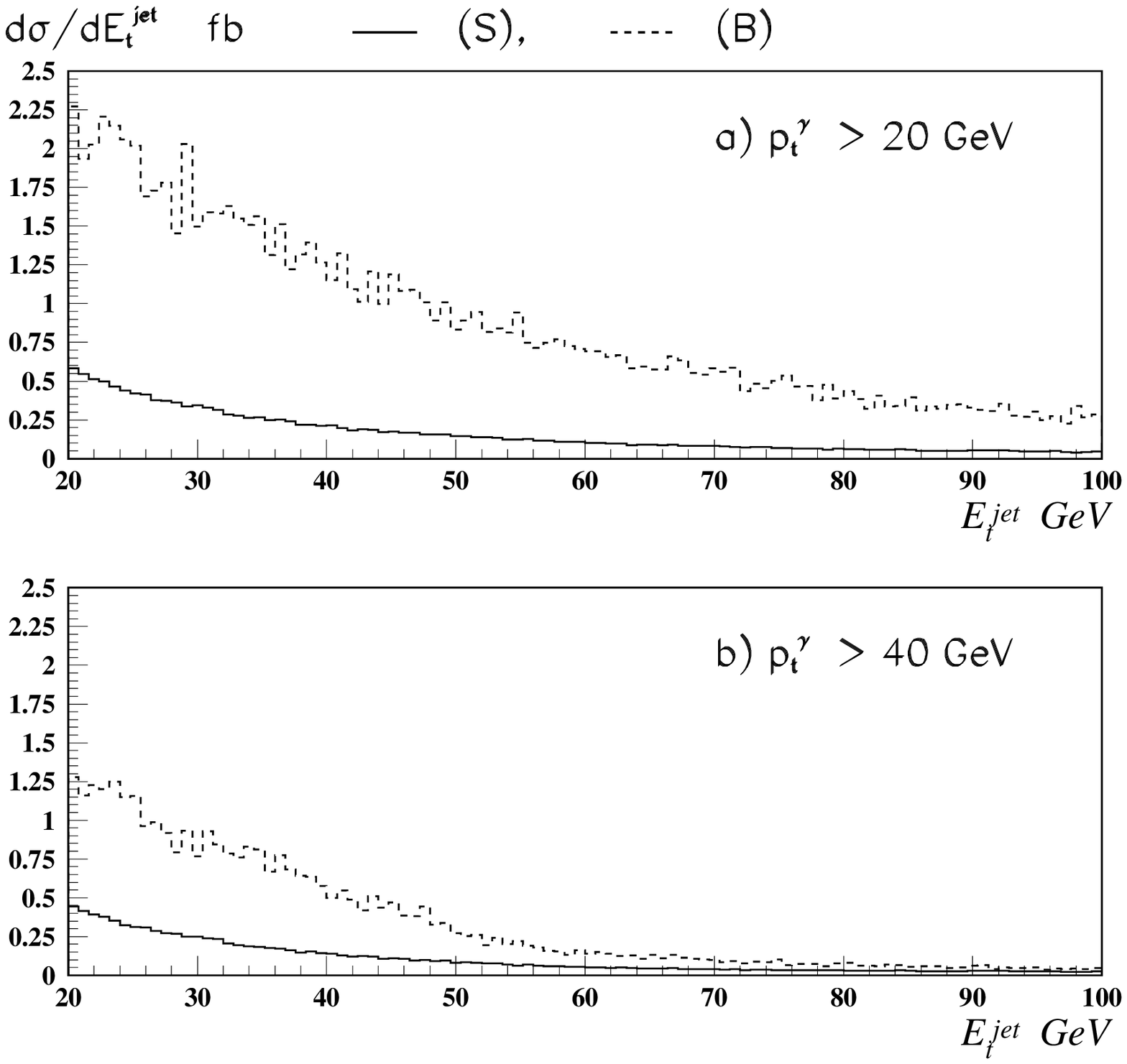}}
\end{picture}
\end{center}
\caption{Distributions in jet transverse energy 
for the main signal (S) and background (B) processes 
$pp\to H+g\to\gamma\gamma +g$ and $pp\to\gamma+\gamma+q$. 
Energy ${\protect \sqrt{s}}=14$ TeV and the Higgs boson mass $M_H=120$ GeV. 
CTEQ4m PDF set and $Q^2=M_H^2+2(E^{jet}_t)^2$ are used.
The basic set of kinematical cuts {\bf (C)} is imposed on other variables.
The $\gamma\gamma$ invariant mass for the background is integrated over the 
range $M_H-\Delta M_{\gamma\gamma} < M_{\gamma\gamma} < M_H +
\Delta M_{\gamma\gamma}$ with $\Delta M_{\gamma\gamma}=1$ GeV. 
\label{fig:aaj-etj}}
\end{figure}

\begin{figure}[hb]
\begin{center}
\unitlength=1cm
\begin{picture}(16,16)
\put(0,0){\epsfxsize=16cm \leavevmode \epsfbox{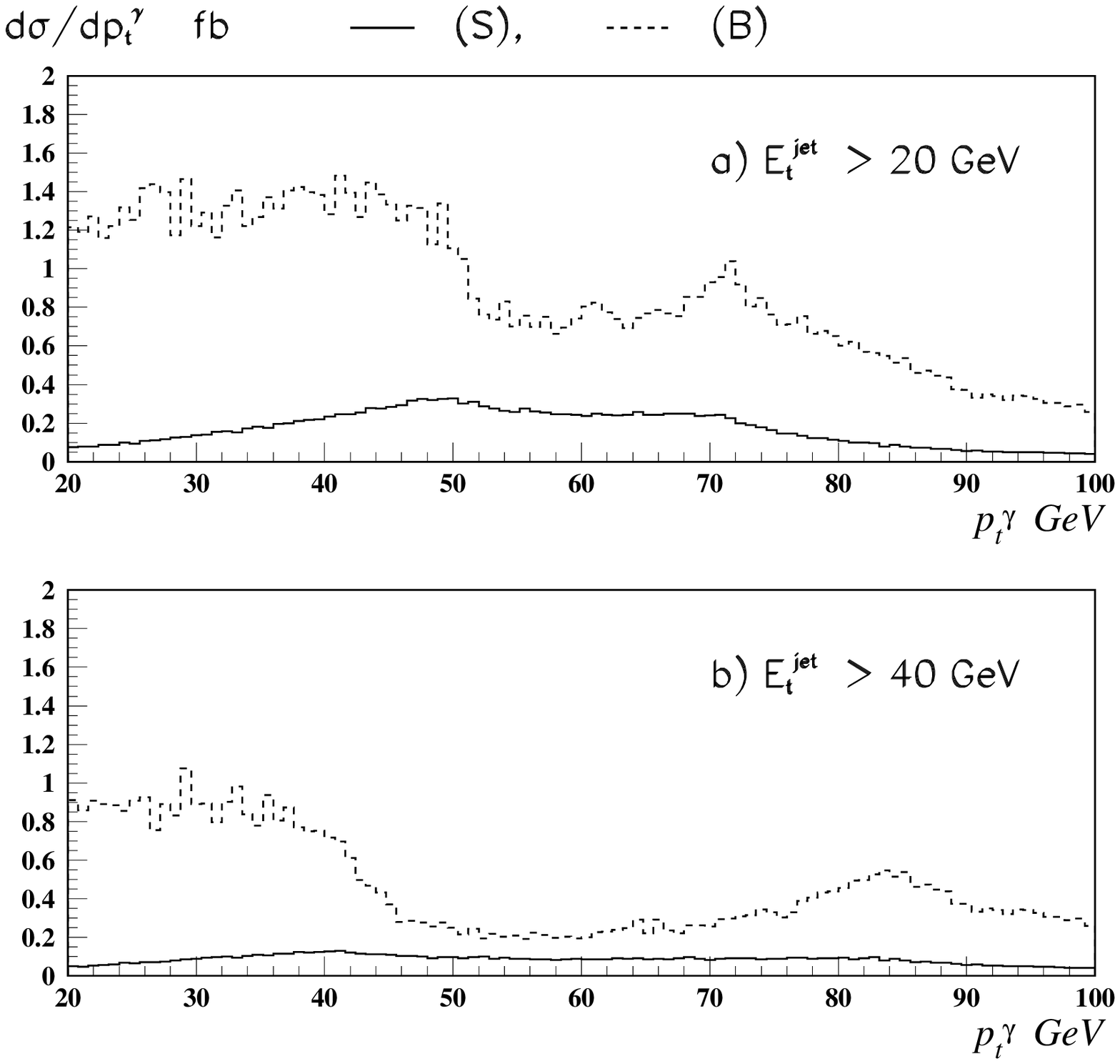}}
\end{picture}
\end{center}
\caption{Distributions in photon transverse momentum 
for the main signal (S) and background (B) processes 
$pp\to H+g\to\gamma\gamma +g$ and $pp\to\gamma+\gamma+q$. 
Energy ${\protect \sqrt{s}}=14$ TeV and the Higgs boson mass $M_H=120$ GeV. 
CTEQ4m PDF set and $Q^2=M_H^2+2(E^{jet}_t)^2$ are used.
The basic set of kinematical cuts {\bf (C)} is imposed on other variables.
The $\gamma\gamma$ invariant mass for the background is integrated over the 
range $M_H-\Delta M_{\gamma\gamma} < M_{\gamma\gamma} < M_H +
\Delta M_{\gamma\gamma}$ with $\Delta M_{\gamma\gamma}=1$ GeV. 
\label{fig:aaj-ptg}}
\end{figure}

The second group of signal subprocesses is made up of electroweak reactions
with Higgs production through $WW$ or $ZZ$ fusion, where one spectator quark
produces the jet with high $E_t$.  Of course, if one discusses the signature
with only one jet one should assume that the set of kinematical cuts applied
{\it vetos} one quark jet in these 
processes.  Here one has also to account for a reaction where the Higgs boson
is produced in association with the $W$- or $Z$-bosons decaying into quark
pairs. In Fig.~\ref{fig:fd_s_WZ} Feynman diagrams for these subprocesses are
shown. The $W/Z$ fusion processes were calculated in \cite{Daw84,KRR84}, while
the processes of $W/Z$ associated production were considered in \cite{e1}. In
total the $W/Z$ fusion and association rates are about 33-40\% of the QCD
signal channels for $M_H=100$-140 GeV. Here we used $Q^2=(M_V/2)^2$ as the
parton factorization scale  for the fusion processes ($V$ denotes $W$ or $Z$),
and $Q^2=(M_V+M_H)^2$ for the associated production.

Note that the process $pp\to H+t+\bar t$ can also contribute to the final state
$\gamma\gamma+jet$. However, our preliminary estimate has shown that the
corresponding rate is very small. So, to  be conservative we do not take into
account this signal channel in our analysis.

In total, we obtain 4.1-6.6 fb total cross section of the signal subprocesses in the 
Higgs mass range $M_H=100$-140 GeV. 

\vskip 0.5cm

Let us now look at the irreducible background. As we already discussed in the
previous section, the subprocess $gq\to\gamma+\gamma+q$ gives the main
background. Another channel, $q\bar q\to\gamma+\gamma+g$, gives about 36\% of
the main $gq$ contribution. Feynman diagrams of these subprocesses are shown
in Fig.~\ref{fig:fd_Birr}.

The one-loop background process $gg\to\gamma+\gamma+g$ exists, 
when the photons are radiated from the quark loop.
Unfortunately this subprocess is not calculated yet. We estimate its cross
section by means of PYTHIA 6.1 simulations. For this purpose one can initialize
PYTHIA subprocess 114, $gg\to\gamma+\gamma$, and switch on gluon bremsstrahlung
from the initial states. This is definitely only one of many other physical
contributions to this process. The result of these simulations shows us that
this background is about 2-4\% of the overall contribution coming from the $gq$
and $q\bar q$ channels. 

In total, the irreducible background amounts to 10.9-19.3 fb{\large /}2GeV for the
Higgs mass range under discussion ($M_H=100-140$ GeV). 

%--------------------------------------------------------------------------
\subsection{Reducible QCD background}

In this section we discuss the results of PYTHIA 6.1 simulations for various
QCD processes which could give a background  due to  radiation of photons from
the fragmentating quarks or gluons. Note that the photon production mechanism
is related in particular to a $\pi^0$-meson (and other neutrals) production --
the energetic $\pi^0$'s could result in a photon pair which will be detected as
a single photon in the electromagnetic calorimeter. Thus, an important
characteristic is the $\pi^0/\gamma$ rejection factor of the calorimeter.  For
example, for CMS detector it is assumed \cite{cmsTP} to be $\sim 3$ from
the possibility to distinguish the two photon clusters. Nevertheless, we do not
take into account this factor in our present analysis.

One kind of reducible background  is coming from the subprocesses  
$$    gq\to\gamma+g+q, \qquad gg\to\gamma+q+\bar q, 
                   \qquad  qq'\to \gamma+q(g)+q'(g),  
$$
in the cases when the final gluon or quark produces an energetic photon and the
jet escapes the detection. The corresponding rates were estimated by the
following algorithm. For definitness let us consider the subprocess 
$gq\to\gamma+g+q$. Firstly, we calculate its cross section by means of CompHEP
imposing the kinematical cuts for a gluon the same as for a photon. We apply
the set {\bf (C)} of kinematical cuts on this 3-body final state and integrate
in the invariant mass of photon and gluon (regarded as a photon) over the
interval of 2 GeV around the central point $M_{\gamma\gamma}=M_H$. The cross
section turns out to be $\sigma\sim 2$ pb ($\sigma'\sim 2.7$ pb when the final
quark is considered as a photon). Then the probability to get a high energy
($>40$ GeV) photon from the fragmentating quark or gluon without further
generation of a detectable jet is estimated as $P(\gamma/q_{veto})\sim 2\cdot
10^{-4}$ and $P(\gamma/g_{veto})\sim 0.3\cdot 10^{-4}$\cite{pcomm}. As a result
we have an estimate of this background as $[\sigma\times
P(\gamma/q_{veto})+\sigma'\times P(\gamma/g_{veto})]\sim$1.1 fb{\large /}2GeV.
Corresponding contributions from the $gg$ and $qq'$ subprocesses are estimated
in the same manner and the result is $\sim$0.5 fb{\large /}2GeV for each
channel. Altogether these channels give a background at the level of $\sim 2.1$
fb{\large /}2GeV.

Another kind of reducible background could come from the subprocesses 
$$gq\to\gamma+q, \qquad q\bar q\to\gamma+g,$$ when the second photon is
produced during the quark or gluon fragmentation but this jet is still
detected. Let us consider the $gq$ channel. As a first step we evaluate with the help
of CompHEP the cross section of this process, applying the $p_t>40$ GeV,
$|\eta_\gamma|<2.5$ and $|\eta_q|<2.5$ cuts, and it turns out to be of order of
2800 pb. For further PYTHIA simulations we switch on a fragmentation in this
process and look for two photons (which can be, in particular, a photon plus
$\pi^0$) satisfying the corresponding cuts from the set {\bf (C)}. We require
also that the invariant mass of this pair should be in the interval
$90<M_{\gamma\gamma}<140$ GeV. In addition we require these photons to be
separated: no charged particles in the cone $\Delta R<0.3$ are produced and the
total energy of neutral particles in the conical area $0.04<\Delta R<0.3$ is
less than 2 GeV. Then we look at these events and try to find, with the help of
the jet finder built in PYTHIA, a jet having proper $E_t$ and  separated from
the generated `good' photons. We generated in total more than $3\cdot 10^6$ events
and no good events were found. As a result we estimate this background  to be
$\le$0.4 fb{\large /}2GeV.  The analogous estimate for the channel $q\bar
q$ is much smaller ($\le$0.01 fb{\large /}2GeV). 

From this latter simulations one can extract a parameter which is useful for
estimate of the next type of reducible background. The probability to get an
energetic photon from the fragmentating 40 GeV jet, separated from
each other so that the jet is still detectable, can be  estimated at a level of
$P(\gamma/jet_{det})\sim 4\cdot 10^{-4}$ \cite{pcomm}. So, this background from
the subprocesses $gq\to\gamma+q$ can be estimated as $2800\mbox{pb}\times
4\cdot 10^{-4}\times \hat P(\gamma,\gamma)$, where $\hat P(\gamma,\gamma)$ is
the unknown probability for two photons (or the photon and $\pi^0$) to be
separated and to have their invariant mass within the interval $M_H-1
GeV<M_{\gamma\gamma}<M_H+1 GeV$. From the above simulations  one obtains $\hat
P(\gamma,\gamma)\le 4\cdot 10^{-5}$.

Third kind of reducible background could comes from the pure QCD
subprocesses of $2\to 2$ type, when both particles in the final state are
gluons or quarks.  We have to discuss these processes in connection with the
probability to get two separated and energetic photons from the fragmentating
quarks and gluons. There are possible contributions from the following
subprocesses\footnote{Note that there is some double counting for the first
type of QCD reducible background analysed in the beginning of this section due
to the photon radiation from the final quarks. We neglect this double counting
in our rather rough estimates. The same remark is valid for the last type of
the QCD background discussed in this section.}
$$ gg\to g(q)+g(\bar q), \qquad  gq\to g+q, \qquad qq'\to q(g)+q'(g). $$
We estimate the corresponding rates by the following formula
$$  \sigma \times P\left(\frac{\gamma}{jet_{det}}\right) 
           \times P\left(\frac{\gamma}{jet'_{veto}}\right)
           \times \hat  P(\gamma,\gamma). $$
Here $\sigma$ stands for the cross section of the subprocess when the
cuts $p_t>40$ GeV and $|\eta_{jet}|<2.5$ are applied.  For the $gg$ channel
$\sigma=1.3\cdot 10^7$ pb. However, this enormous cross section is reduced to
the insignificant level of $\le 1.4\cdot 10^{-3}$ fb by probabilities to get
the proper photons. For other channels $\sigma$ has values much smaller and we
have rates of $\le 10^{-3}$ fb for the $gq$ and of $\le 10^{-4}$ fb for
the $qq'$ ones. In total, this type of background gives a very small 
contribution of $\sim 0.0025$ fb{\large /}2GeV.

Finally we consider reactions of the $2\to 3$ type when all three particles in
the final state are gluons or quarks. At the parton level these processes can
be  derived from the processes discussed in the previous paragraph taking into
account gluon radiation. So, as in the previous case, there are three subgroups
of such processes with the $gg$, $gq$ and $qq'$ initial states. Here we have to
account for events where two separated and energetic photons are produced by
fragmentating quarks or gluons which escape detection.  Let us consider the 
subprocess $gg\to g+g+g$. If we consider two of final gluons as photons and
apply our basic set {\bf (C)} of kinematical cuts the cross section is of order
of $5.8\cdot 10^3$ pb{\large /}2GeV. To get the necessary estimate we use the
formula 
$$ \sigma \times P\left(\frac{\gamma}{jet_{veto}}\right) 
           \times P\left(\frac{\gamma}{jet'_{veto}}\right).$$

In the $gg$ case we have to multiply this formula by a combinatorial factor 3 due
to three variants to produce two photons from different gluons. We
neglect the possibility when two photons are produced from the same gluon jet
while the other gluon jet escapes detection. So, this background is estimated
at the level of $\sim 0.016$ fb{\large /}2GeV. The analogous estimates for
other subprocesses of this type give $0.018$ fb{\large /}2GeV for $gq$ and
$0.002$ fb{\large /}2GeV for the $qq'$ channels. In total, this type of QCD
background gives rather small contribution of  $\sim 0.036$ fb{\large /}2GeV.

Altogether four types of QCD reducible background processes contribute with
a rate of $\le 2.2$ fb{\large /}2GeV. This is not a negligible background and
one has to take it into account analysing  the reaction
$pp\to\gamma\gamma+jet$. 

Now we  can conclude that the QCD reducible background, connected with the
probability to get a photon from the fragmentating gluons or quarks, is
potentially dangerous, especially if we remember various uncertainties in our
estimates.  Nevertheless, this background turns out to be less than 20\% of
the irreducible background.

%-------------------------------------------------------------------------
\subsection{\boldmath Summary }

From Table~\ref{tab:aaj-all} one can see, first of all, that  with an
integrated luminosity of 10 fb$^{-1}$  this channel can give 40-70 signal
events  with a number of background events only by a factor of 2-3 higher. This
result was obtained for the basic set of kinematical cuts {\bf (C)} (see
Section 4.2) and for resolution determined mass interval $M_H\pm 1$ GeV in the
two-photon invariant mass $M_{\gamma\gamma}$. The significances are
$N_S/\sqrt{N_B}\sim$4.0, 5.3 and 4.1 for $M_H=100$, 120 and 140 GeV,
respectively, showing good prospects for discovery of the Higgs boson in
this reaction at low LHC luminosity, when total statistics collected is $\sim
30$ fb$^{-1}$.  These results also mean that each year of LHC operation at high
luminosity of $10^{34}\mbox{cm}^{-2}\mbox{s}^{-1}$ will give hundreds of events
with high $p_t$ Higgs bosons associated with high $p_t$ jet in central
detector. The signal significance will be of order of 15 in this case.

Most improtant advantage of the reaction $pp\to\gamma\gamma+jet$ in comparison
with the 'standard' channel $pp\to\gamma\gamma$ is significant improvement of
the {\it Signal/Background} ratio. Let us remind that in $pp\to\gamma\gamma$
this ratio is $\sim 1/15$ (calculated in the LO approximation and with 2 GeV
photon invariant mass intevral for the background, as in our case). For the
reaction $pp\to\gamma\gamma+jet$ the {\it Signal/Background} ratio is $\sim
1/2-1/3$. Signal significance $N_S/\sqrt{N_B}$ is approximately the same for
both channels. 

However, we should stress that in present analysis we did not account for
various factors which can change the Higgs significance considerably. Let us
review these factors and start from those which should help to improve the
values obtained for the Higgs signal significance.

First, the results presented in Table~\ref{tab:aaj-all} for hard subprocesses
have been obtained in leading orders $\alpha_s^3$ for the QCD signal processes
and $\alpha_s$ for the background. As we have mentioned in Section 4.1 the 
next-to-leading QCD corrections are unknown for these
subprocesses.\footnote{Some progress has appeared recently for the signal, see
Ref.~\cite{Carl}.} Notice that the NLO corrections to the hard subprocesses in
the $pp\to\gamma\gamma+X$ inclusive reaction increase the Higgs production
cross section by about 60\% (see \cite{SDGZ} and the references therein). Of
course, it is very probable that the NLO corrections enhance somehow the rates
of background processes as well. However, due to a lower strong coupling order
of the background processes  one can assume their QCD corrections to be smaller
than for the signal. 

There are possible ways  to improve the significance of the $\gamma\gamma+jet$
channel by applying weaker jet cuts. In particular, if the detection of jet is
possible at large rapidities, up to $|\eta_{jet}|\sim 4$ (involving the very
forward hadron calorimeter), the significance will increase by $\sim 14$\%.
Furthermore, if the  determination of jet is reliable with $E^{jet}_t>30$ GeV
then the significance will increase by $\sim 8$\% in comparison with the case
of basic kinematical cuts {\bf (C)}.

Let us remember now that for the CMS PbWO$_4$ electromagnetic calorimeter the
two-photon invariant mass resolution is expected to be significantly better 
than 1~GeV in low luminosity running. For example,  in \cite{cmsTP} the
following two-photon mass resolutions were considered for the reaction
$pp\to\gamma\gamma+X$ at low luminosity: $\sigma^L_m=540$ and 600 MeV for
$M_H=110$ and 130 GeV, respectively. Surely, in this case the Higgs
significance will increase noticeably compared to the estimates made earlier in
this paper.

However, there are also many factors reducing signal significance. One of them
is that the one-loop subprocess $gg\to\gamma+\gamma+g$ contributing to the
background, unfortunately, is not still calculated -- its exact matrix element
is unknown. Our estimate of its rate is rather small, $\sim$ 0.2-0.8 fb{\large
/}2GeV. However, these numbers have been obtained by means of PYTHIA
simulations for the process $gg\to\gamma+\gamma$ where the final gluon is
radiating from the initial state only. We expect the rate of this background to
be higher due to other mechanisms of the final gluon radiation.

Furthermore, the reducible QCD background is potentially dangerous as a rule
at hadron colliders. We have estimated this background at the level of $\le$2.2
fb{\large /}2GeV. Of course, more accurate evaluation of this background would
be desirable taking into account the detailed simulation of the ATLAS and CMS
detectors. Due to this discussion one can remember that similar analysis for
the inclusive $pp\to\gamma\gamma+X$ reaction showed \cite{cmsTP,atlasTDR} that
the reducible QCD background can be suppressed at the level of $<10$\% of the
irreducible bias.

The effeciency of photons detection was not taken into account in our
estimates. In our case the photons have large transverse momenta
$p_t^\gamma>40$ GeV and radiated with small rapidities $|\eta_\gamma|<2.5$. So
one can hope that the photon detection efficiency will be high enough. At
present time for both ATLAS and CMS detectors this efficiency is estimated at
the level $\approx 0.7-0.8$. Experimental efficiency of hard jet detection was
also not taken into account.  

Finally, combining all these factors one can hope that they could at least 
compensate  each other, and our general conclusion about a possibility to
discover the light Higgs boson during the LHC operation at low luminosity is
still kept.

%#############################################################
\section*{Conclusions}

The channel $\gamma\gamma+jet$ gives very promising discovery possibilities for
the Higgs boson with a mass of 100-140 GeV during the LHC operation
at low luminosity of $\sim 10^{33}$cm$^{-2}$s$^{-1}$. With an integrated
luminosity of 30 fb$^{-1}$ 120-200 signal events could be observed with 330-600
background events per 2~GeV interval of two photon invariant mass, showing a
signal significance of $\sim 7-8.5$. For high LHC luminosity of 
$10^{34}$cm$^{-2}$s$^{-1}$  hundreds of Higgs bosons will be observed with a
significance of order of 15 for one year data accumulation. The reducible QCD
background is at the level of $\le$20\% of the irreducible one. The
next-to-leading QCD corrections, not calculated at present time, could enhance
the signal more than the background. Higher two-photon invariant mass
resolution that is now under discussion for the CMS PbWO$_4$ calorimeter (see
\cite{cmsTP}, $\sigma^L_m\sim 500-600$ MeV at $M_H=110-130$ GeV) could increase
the significance of signal. On the other hand, the efficiencies of
photon and jet detection that we did not take into account decrease the signal
significance. These factors (as well as others discussed above) work in the
opposite directions and it is probable that they compensate each other.

%--------------------------------------------- 
\section*{Acknowledgements} 

We are grateful to D.~Denegri and N.~Stepanov for attracting our attention to
the signatures with high $p_t$ jets.  We thank very much them and S.~Abdullin
for the fruitful collaboration as well as E.~Boos, M.~Mangano and A.~Pukhov for
many useful discussions. We are grateful to D.~Kovalenko for his collaboration
in implementing the loop vertices in CompHEP.  Our simulations by PYTHIA would
have been impossible without the help of S.~Abdullin and N.~Stepanov. We are
also indebted to them and I.~Goloutvine, A.~Nikitenko and R.~Kinnunen for the
interesting discussions about the CMS detector.

Our participation in the CMS Collaboration has been realized due to the kind
promotion of V.~Matveev and I.~Goloutvine and the support of the Russian 
Ministry of Science and Technologies. This work was partially supported also by
the European Association INTAS (grant 93-1180ext) and the Russian Foundation 
for Basic Research (grants 96-02-18635 and 96-02-19773).

We express our sincere gratitude to the CMS Collaboration and the CERN PPE
Division for the hospitality and kind assistance during our stay at CERN.

%\newpage
%========================== Bibliography ==========================

\clearpage
%==========================  Tables ===============================
\section*{Tables}

\begin{table}[hb]
\vspace*{0.5cm}
\large a)
\begin{center}
{\large
\begin{tabular}{|c|c|r|r|r|r|r|r|r|r|r|}
\hline
\multicolumn{2}{|c|}{\raisebox{0ex}[3ex][2ex]{$|\eta^j| <$}} 
& 1.00& 1.50& 2.00& 2.40& 3.00& 3.50& 4.00& 4.50& 5.00\\
\hline
$\sigma$  & \raisebox{0ex}[3ex][2ex]{\bf S}  
& 2.16& 3.10& 3.84& 4.32& 4.79& 5.01& 5.12& 5.14& 5.16\\
\cline{2-11} 
fb        & \raisebox{0ex}[3ex][2ex]{\bf B} {\Large /}2 GeV  
& 6.48& 9.16&11.04& 11.89& 12.64& 12.76& 12.80&
12.80& 12.82\\
\hline
\multicolumn{2}{|c|}{\raisebox{0ex}[3ex][2ex]
             {$\frac{\sigma_S}{\sqrt{\sigma_B}}$}}             
& 0.85& 1.03& 1.15& 1.25& 1.35& 1.40& 1.43& 1.44& 1.44\\
\hline
\multicolumn{2}{|c|}{\raisebox{0ex}[3ex][2ex]
             {$\frac{\sigma_S}{\sigma_B}$}}
& 0.33& 0.34& 0.35& 0.36& 0.38& 0.39& 0.40& 0.40& 0.41\\
\hline
\end{tabular}
}
\end{center}

\vspace*{0.5cm}

\large b)
\begin{center}
{\large
\begin{tabular}{|c|c|r|r|r|r|r|r|r|}
\hline
\multicolumn{2}{|c|}{\raisebox{0ex}[3ex][0ex]{$E^{jet}_t >$}} 
                               & 20  & 25  & 30  & 35  & 40  & 45  & 50   \\
\multicolumn{2}{|c|}{GeV}      &     &     &     &     &     &     &      \\ 
\hline
$\sigma$  & \raisebox{0ex}[3ex][2ex]{\bf S} 
                               & 9.25& 7.43& 6.05& 5.07& 4.32 & 3.73& 3.27 \\
\cline{2-9} 
fb        & \raisebox{0ex}[3ex][2ex]{\bf B} {\Large /}2 GeV                 
                               &30.86&24.72& 19.62& 15.32&11.89 & 9.22& 7.24 \\
\hline
\multicolumn{2}{|c|}{\raisebox{0ex}[3ex][2ex]
         {$\frac{\sigma_S}{\sqrt{\sigma_B}}$}}
                               &1.66 & 1.49& 1.36 &1.29 & 1.25& 1.23 & 1.22 \\
\hline
\multicolumn{2}{|c|}{\raisebox{0ex}[3ex][2ex]
         {$\frac{\sigma_S}{\sigma_B}$}}
                               & 0.30& 0.30& 0.31& 0.33& 0.36& 0.41& 0.45 \\
\hline
\end{tabular}
}
\end{center}
\caption{Cross sections of the main signal ({\bf S}) and background ({\bf B}) 
processes $pp\to H+g\to\gamma\gamma +g$ and $pp\to \gamma+\gamma+q$, as a
function of the jet a) rapidity cuts and b) transverse energy. 
The basic set {\bf (C)} of cuts is imposed on
other variables. $M_H=120$ GeV. PDF set CTEQ4m is used. All subprocesses are
evaluated in QCD leading order and the 2nd order running $\alpha_s$ with
normalization $\alpha_s(M_Z)=0.118$. $Q^2=M_H^2+2(E^{jet}_t)^2$ is taken as the 
QCD scale for running $\alpha_s$  and as the parton factorization scale. 
The $\gamma\gamma$ invariant mass for the background is integrated over the 
range $M_H-\Delta M_{\gamma\gamma} < M_{\gamma\gamma} < M_H +
\Delta M_{\gamma\gamma}$ with $\Delta M_{\gamma\gamma}=1$ GeV. 
\label{tab:aaj-jcuts} }
\end{table}

\clearpage

\begin{table}[hb]
\begin{center}
\large
\begin{tabular}{|c|c|c|c|c|}
\cline{2-5} 
\multicolumn{1}{c|}{ }&\raisebox{0ex}[3ex][0ex]{Hard subprocess} &
\multicolumn{3}{c|}{\large $\sigma$, fb}\\
\cline{3-5}
\multicolumn{1}{c|}{ }&  & $M_H=100$ GeV & 120 GeV & 140 GeV \\
\hline
   &  $gg\to H+g$                          & 2.55  & 4.32  & 3.75  \\
{\large \bf S}   &  $gq\to H+q$            & 0.39  & 0.59  & 0.50  \\
 &  $q\bar q\to H+g$                       &0.006  & 0.007 & 0.006 \\
\hline
&  $qq'\to H+q+q'$                         & 0.88  & 1.33  & 1.16  \\
{\large \bf S}  & $q\bar q'\to H+W$        & 0.25  & 0.31  & 0.19  \\
($jet_{det}+jet_{veto}$)& $q\bar q\to H+Z$ 
                                           & 0.071 & 0.087 & 0.054 \\
\hline \hline
&\raisebox{0ex}[3ex][0ex]{$gq\to\gamma+\gamma+q$}
                                           & 7.83  & 11.89 & 14.86  \\
{\large\bf B} {\Large /} 2 GeV
   &  $q\bar q\to\gamma+\gamma+g$          & 2.84  & 3.50  & 3.64  \\
   &  $gg\to\gamma+\gamma+g$ $^{(*}$ & $\sim 0.2$ &$\sim 0.5$ & $\sim 0.8$\\
\hline 
{\large \bf B} {\Large /} 2 GeV  
& \raisebox{0ex}[3ex][0ex]{ $gq\to\gamma+g+q$ $^{(*}$ }
                                           &\multicolumn{3}{c|}{$\sim 1.1$}\\
$\left(\frac{\gamma}{jet_{veto}}\right)$ &
 $gg\to\gamma+q+\bar q$ $^{(*}$            &\multicolumn{3}{c|}{$\sim 0.5$} \\
 &  $qq'\to\gamma+q+q'$                    &\multicolumn{3}{c|}{$\sim 0.5$}\\
\hline
\raisebox{0ex}[3ex][0ex]{\large \bf B} {\Large /} 2 GeV  &
  $gq\to\gamma+q$ $^{(*}$                  &\multicolumn{3}{c|}{$\le 0.4$} \\
\raisebox{0ex}[0ex][2ex]{$\left(\frac{\gamma}{jet_{det}}\right)$} &
  $q\bar q\to\gamma+g$ $^{(*}$             &\multicolumn{3}{c|}{$\le 0.001$}  \\
\hline 
{\large \bf B} {\Large /} 2 GeV   
& \raisebox{0ex}[3ex][0ex]{ $gg\to g+g$ $^{(*}$ }
                                           &\multicolumn{3}{c|}{$\le 0.0014$} \\
$\left(\frac{\gamma}{jet_{det}}+\frac{\gamma}{jet_{veto}}\right)$ 
& \raisebox{0ex}[0ex][2ex]{$gq\to g+q$ $^{(*}$ }  
                                           &\multicolumn{3}{c|}{$\le 0.001$}\\
& \raisebox{0ex}[0ex][2ex]{$qq'\to q+q'$ $^{(*}$ }  
                                           &\multicolumn{3}{c|}{$\ll 0.001$}\\
\hline
\hline 
{\large \bf B} {\Large /} 2 GeV   
& \raisebox{0ex}[3ex][0ex]{$gg\to g+g+g$ $^{(*}$ }
                                           &\multicolumn{3}{c|}{$\sim 0.016$} \\
$\left(\frac{\gamma}{jet_{veto}}+\frac{\gamma}{jet_{veto}}\right)$ 
& \raisebox{0ex}[0ex][2ex]{$gq\to g+g+q$ $^{(*}$ }  
                                           &\multicolumn{3}{c|}{$\sim 0.018$}\\
& \raisebox{0ex}[0ex][2ex]{$qq'\to g+jet+jet'$ $^{(*}$ }  
                                           &\multicolumn{3}{c|}{$\sim 0.002$}\\
\hline
\end{tabular}
\end{center}
\caption{Summary for the $pp\to\gamma\gamma+jet$ reaction.
Contributions of different subprocesses are shown
for the basic set of kinematical cuts {\bf (C)} (see Section 4.2). PDF set
CTEQ4m is used. All subprocesses are evaluated in QCD leading order and the 
2nd order running $\alpha_s$ with normalization $\alpha_s(M_Z)=0.118$.
$Q^2=M_H^2+2(E^{jet}_t)^2$ was taken as the QCD scale for running $\alpha_s$ 
and as the parton factorization scale in hard QCD subprocesses. The values
$Q^2=(M_V/2)^2$ and $Q^2=(M_V+M_H)^2$, where $V=(W,Z)$, are used for the parton
factorization scale in the $W/Z$ fusion and $W/Z$ associated Higgs signal 
procceses, correspondingly. The $\gamma\gamma$ invariant mass for the background
is integrated over the range $M_H-\Delta M_{\gamma\gamma} < M_{\gamma\gamma} 
< M_H +\Delta M_{\gamma\gamma}$ with $\Delta M_{\gamma\gamma}=1$ GeV. 
Asterisk `$^*$' marks the results of the PYTHIA 6.1 simulations.  
Other results are obtained with the help of CompHEP package. 
\label{tab:aaj-all} }
\end{table}

\clearpage
%============================ Figures =============================
\section*{Figures}

\vskip 2cm

\clearpage

%\begin{figure}[hb]
%\begin{center}
%\unitlength=1cm
%\begin{picture}(16,16)
%\put(0,0){\epsfxsize=16cm \leavevmode \epsfbox{ygammas.eps}}
%\end{picture}
%\end{center}
%\caption{Distributions in two-photon rapidity 
%for the main signal (S) and background (B) processes 
%$pp\to H+g\to\gamma\gamma +g$ and $pp\to\gamma+\gamma+q$. 
%Energy ${\protect \sqrt{s}}=14$ TeV and the Higgs boson mass $M_H=120$ GeV. 
%CTEQ4m PDF set and $Q^2=M_H^2+2(E^{jet}_t)^2$ are used.
%The basic set of kinematical cuts {\bf (C)} is imposed.
%The $\gamma\gamma$ invariant mass for the background is integrated over the 
%range $M_H-\Delta M_{\gamma\gamma} < M_{\gamma\gamma} < M_H +
%\Delta M_{\gamma\gamma}$ with $\Delta M_{\gamma\gamma}=1$ GeV. 
%\label{fig:aaj-etagdi}}
%\end{figure}

\end{document}